\documentclass[preprint,double-spaced,floatfix,showpacs,letterpaper]{revtex4}
\usepackage{graphicx}
\usepackage{bm}

\begin{document}
\topmargin-1.3cm

\title{Role of dipolar correlations in the IR spectra of water and ice}

\author{Wei Chen$^{1}$, Manu Sharma$^{2}$, Raffaele Resta$^{3,4}$, Giulia Galli$^{2}$, Roberto Car$^{1,5,6}$}
\affiliation{
$^1$Department of Physics, Princeton University, Princeton, NJ 08544, USA\\
$^2$Department of Chemistry, University of California, Davis, CA 95616, USA\\
$^3$INFM DEMOCRITOS National Simulation Center, via Beirut 2, 34014 Trieste, Italy\\
$^4$Dipartimento di Fisica Teorica, Universit\`a di Trieste, Strada Costiera 11, 34014 Trieste, Italy\\
$^5$Department of Chemistry, Princeton University, Princeton, NJ 08544, USA\\
$^6$Princeton Institute for the Science and Technology of Materials, Princeton University, Princeton, NJ 08544, USA}

\date{\today}

\begin{abstract}
We report simulated infrared spectra of deuterated water and ice using Car-Parrinello molecular dynamics with maximally localized Wannier functions. Experimental features are accurately reproduced within the harmonic approximation. By decomposing the lineshapes in terms of intra and intermolecular dipole correlation functions we find that short-range intermolecular dynamic charge fluctuations associated to hydrogen bonds are prominent over the entire spectral range. Our analysis reveals the origin of several spectral features and identifies network bending modes in the far IR range.
\end{abstract}

\pacs{36.20.Ng, 71.15.Pd, 32.10.Dk, 78.20.Ci}

\maketitle

\section{\label{sec1}Introduction}
It is well known that correlations among the dipole moments of hydrogen bonded molecules greatly enhance the {\em static} dielectric response of water and ice (see e.g. Ref.~\cite{Sharma07} and references therein). However, the role played by dipolar correlations in the {\em dynamic} dielectric response of hydrogen bonded systems is not well understood, although it has often been speculated that intermolecular couplings affect significantly infrared (IR) and Raman spectra~\cite{Whalley77,Rice83,Buch99,Guillot91,Silvestrelli97}. A difficulty in modeling these spectra is that the dynamic dielectric susceptibility depends on the adiabatic response of the electrons to nuclear dynamics. Another difficulty, in principle even greater, is that nuclear quantum effects cannot be discarded {\em a priori} in water systems at, or below, room temperature, and it is beyond current theoretical and computational capabilities to fully account for dynamic quantum effects in systems of this complexity. Recent studies have shown, however, that the IR absorption lineshapes of liquid water and ice are overall reproduced remarkably well by {\em ab initio} molecular dynamics (AIMD) simulations, which treat nuclear dynamics classically but include explicitly, and accurately, the quantum adiabatic response of the electrons ~\cite{Sharma05,Iftimie05}. 

In this paper we adopt the AIMD framework and use maximally localized Wannier functions (MLWF) to separate intra and inter molecular contributions to the dynamic dipolar correlations in liquid water and ice. We find that hydrogen bonds (H-bonds) make short-range intermolecular fluctuations of the electronic charge as important as the intramolecular fluctuations over the entire spectral range. Our analysis sheds new light into the origin of several spectral features and provides an unambiguous identification of network bending modes in the far IR region~\cite{Zelsmann95}.

The IR absorption coefficient per unit length $\alpha(\omega)$ is related to the refractive index $n(\omega)$ and the imaginary part of the dielectric constant $\epsilon''(\omega)$ by $\alpha(\omega)n(\omega)=(\omega/c)\epsilon''(\omega)$. Within linear response theory, $\alpha(\omega)$ is given by the power spectrum of the time correlation function of the total dipole operator~\cite{McQuarrie00}. Following common practice, we approximate the quantum time correlation function with the classical one, i.e. with $\langle\bm{M}(0)\bm{M}(t)\rangle$, where $\bm{M}$ is the total dipole moment in the simulation cell and the brackets indicate classical ensemble average. Since there is only one classical correlation function while the quantum time correlation function can be expressed in several equivalent ways, this substitution leads to formulae for the IR absorption coefficient characterized by different prefactors known as quantum correction factors~\cite{Ramirez04}. We adopt here the so-called harmonic approximation (HA) which follows by replacing the Kubo-transformed quantum correlation function with the classical one, leading to: 
\begin{equation}
\alpha(\omega)n(\omega) = \frac{2\pi\omega^2\beta}{3cV} \int_{-\infty}^\infty dt e^{-i\omega t}
\langle\bm{M}(0)\bm{M}(t)\rangle \label{lineshapeHA}
\end{equation}
where $\beta=(k_B T)^{-1}$ is the inverse temperature. In the harmonic regime HA is exact~\cite{Bader94}. Ram{\'i}rez {\em et al}.~\cite{Ramirez04} showed that HA is the only correction factor that satisfies the fluctuation-dissipation theorem in addition to detailed balance. The same authors also found that HA performs better than the other quantum correction factors for one dimensional anharmonic potentials that model different H-bond scenarios.  

\section{\label{sec2}Computational details}
In this work, we compute the IR spectra of D$_2$O ice and water using a Car-Parrinello (CP) scheme~\cite{Car85}, in which maximally localized Wannier functions (MLWFs)~\cite{Marzari97} represent ``on the fly'' the electronic wave functions~\cite{Sharma03}. In the simulation of water (ice), we use a periodically repeated cubic (orthorhombic) cell with 64 (96) D$_2$O molecules at the density of 1.1 (1.0) g/ml. For ice we use a proton disordered ice Ih configuration generated as in Ref.~\cite{Sharma07}. We adopt norm-conserving pseudopotentials~\cite{Troullier91} and the DFT Perdew-Burke-Ernzerhof (PBE) functional for exchange and correlation~\cite{Perdew96}, using a plane-wave cutoff of 60 Ry for water and of 85 Ry for ice~\cite{Details_Cutoff}. We use an integration time step of 7 a.u. (0.17 fs) and a fictitious electron mass of 350 a.u. in the CP equations~\cite{Car85,Sharma05,Sharma07}. The temperature is controlled by coupling the nuclei to a single Nos{\'e} thermostat~\cite{Blochl92} with mass $Q_{N}=1\times10^{6}$. In this way the average temperature of the system is set to 330 K in water and to 268 K in ice.

At each time step we assign a dipole $\bm{\mu}_i=\bm{r}_{D_1} ^i + \bm{r}_{D_2} ^i + 6\bm{r}_{O} ^i - 2\sum_{l=1,4}\bm{r}_{W_l} ^i$
to the $i$-th molecule in the cell. Here $\bm{r}_X ^i$ are the positions of the nuclei ($X$ = D$_1$, D$_2$, O) and of the four MLWF centers ($X = W_l$, with $l = 1, 4$) associated to the eight valence electrons of a water molecule~\cite{Silvestrelli99,Sharma05,Sharma07}. The $i$-th dipole is conventionally located at the molecular center of mass $\bm{r}^i$, and depends on the local environment. The total dipole moment is $\bm{M}=\sum_{i=1,N} \bm{\mu}_i$ ~\cite{Pasquarello03,Resta94} and the IR spectra are computed with Eq.~(\ref{lineshapeHA}). The time correlation function is averaged over an equilibrated MD trajectory of 10 ps in ice and of 23 ps in water. A Gaussian window function~\cite{Harris1978} is used in the discrete Fourier transform.

\section{\label{sec3}Results and discussion}
\subsection{\label{sec31}Calculated IR spectra}
\begin{figure}[htp]
   \resizebox{\columnwidth}{!}{\includegraphics{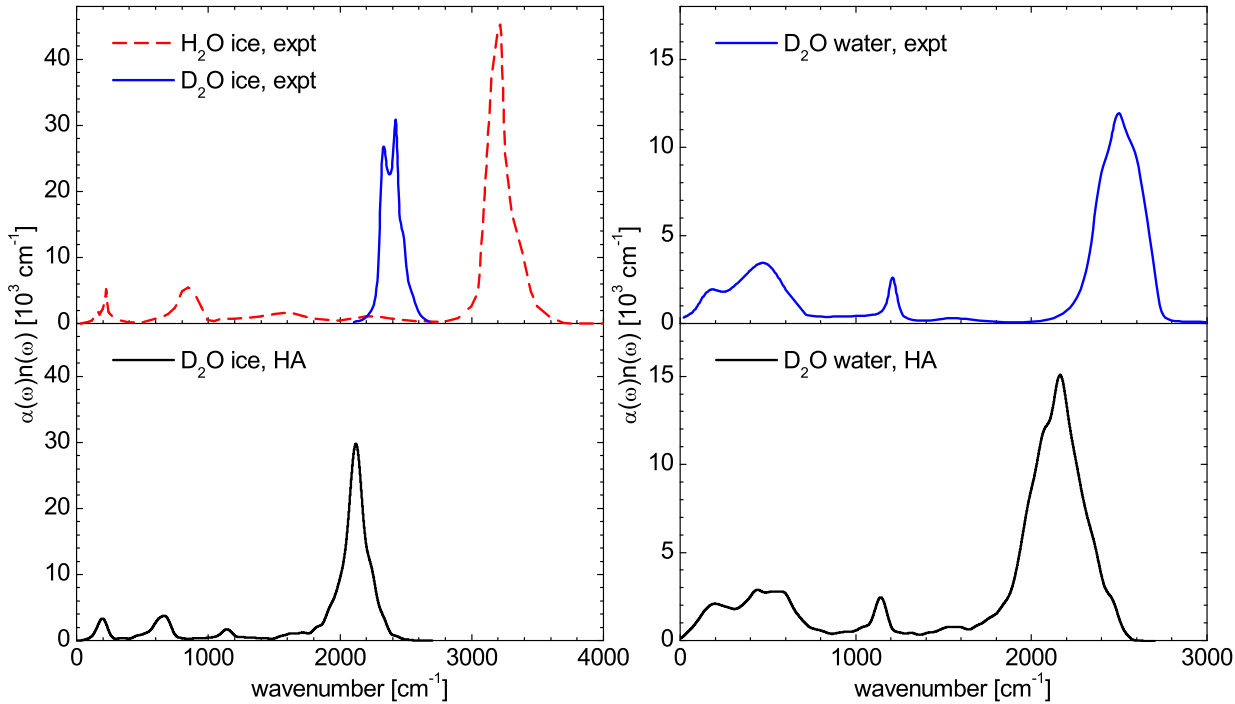}}    
   \caption{ 
(color online). Experimental (upper panels) and calculated IR spectra (lower panels). Experimental data are from Ref.~\cite{Bertie69} for H$_2$O ice at 100K, and from Ref.~\cite{Bergren78} for D$_2$O ice at 150K. D$_2$O water data are from Ref.~\cite{Bertie89} at 295K and from Ref.~\cite{Zelsmann95} at 293K.  Experimental data for D$_2$O ice are not available over the entire range.}
   \label{ir_expt}
\end{figure}

The calculated spectra are compared with experiment in Fig.~\ref{ir_expt}~\cite{Details_smooth}.
As in previous HA calculations~\cite{Sharma05,Iftimie05} the overall agreement between theory and experiment is good. For ice, the positions of the band maxima corresponding to H-bond stretching, libration, bending, combination, and OD stretching are: 200 (222), 660 (640), 1140 (1210), 1660 (1650), and 2120 (2425) cm$^{-1}$ respectively (wavenumbers in parentheses are experimental values from Ref.~\cite{Eisenberg69}). For water, the corresponding values are: 200 (187), 510 (505), 1140 (1215), 1550 (1555), and 2160 (2450) cm$^{-1}$ respectively. The details of the experimental features are well reproduced, e.g. the asymmetric shape of the OD stretching bands, the skewed ice libration band, and even the small combination bands. The good performance of the harmonic approximation may appear somewhat surprising given the large shifts and broadenings of the vibrational frequencies of an isolated molecule that arise in condensed phase. Furthermore, significant anharmonicity has been detected in the excited hydrogen stretching modes in liquid water~\cite{Bakker2002}. IR absorption, however, probes equilibrium properties which are dominated by the vibrational ground-state in the hydrogen stretching region. Then, the overall similarity between calculated and observed spectra suggests that the main effects of anharmonicity are sufficiently well captured by classical dynamics.

The largest discrepancy between simulation and experiment occurs in the hydrogen stretching modes, which are redshifted compared to experiment by $\sim$300 cm$^{-1}$. Part of this error originates from our choice of the mass $\mu$ for the fictitious dynamics of the electrons. We can quantify this effect by calculating the IR stretching band in ice from Born-Oppenheimer (BO) molecular dynamics simulations, in which the electrons are kept in the instantaneous ground-state by minimizing the energy functional at each time step. In CP simulations this condition is enforced dynamically and the outcome depends on the choice of $\mu$. BO calculations show that a redshift of up to $\sim$80 cm$^{-1}$ in the IR stretching band of ice can be attributed to our choice of $\mu$~\cite{Details_mu}. However, even when the BO separation is strictly enforced the IR stretching band in ice is more than 200 cm$^{-1}$ below experiment. The adopted DFT approximation is a likely cause of this error. For instance, in the H$_2$O water monomer, PBE yields stretching frequencies that are $\sim$140 cm$^{-1}$ lower than the corresponding experimental values~\cite{Xu2004}. Taking the isotopic mass effect into account we should expect a redshift of $\sim$100 cm$^{-1}$ for the stretching band of D$_2$O in gas phase. The actual effect that we observe in condensed phase, $\sim$200 cm$^{-1}$, is larger than that and is consistent with the known H-bond over binding present in PBE water. Other sources of inaccuracy in the calculated spectra include the finite basis set, the cell size and the effect of temperature. Quantum anharmonic corrections are beyond our approach.

The similarity between the spectra of solid and liquid water is consistent with the standard picture of a tetrahedral, ice-like local structure of the liquid. With more disorder in the liquid phase, modes close in frequency overlap with each other, leading to broader bands than in the solid. Fluctuating local fields and coupling between neighboring molecules cause additional broadening. The relative importance of local fields and intermolecular couplings can be quantified in terms of intra and inter spectral contributions.

\subsection{\label{sec32}Intra vs. inter contribution}
\begin{figure*}[htp] 
\begin{center}
\includegraphics [width=0.91\textwidth, clip] {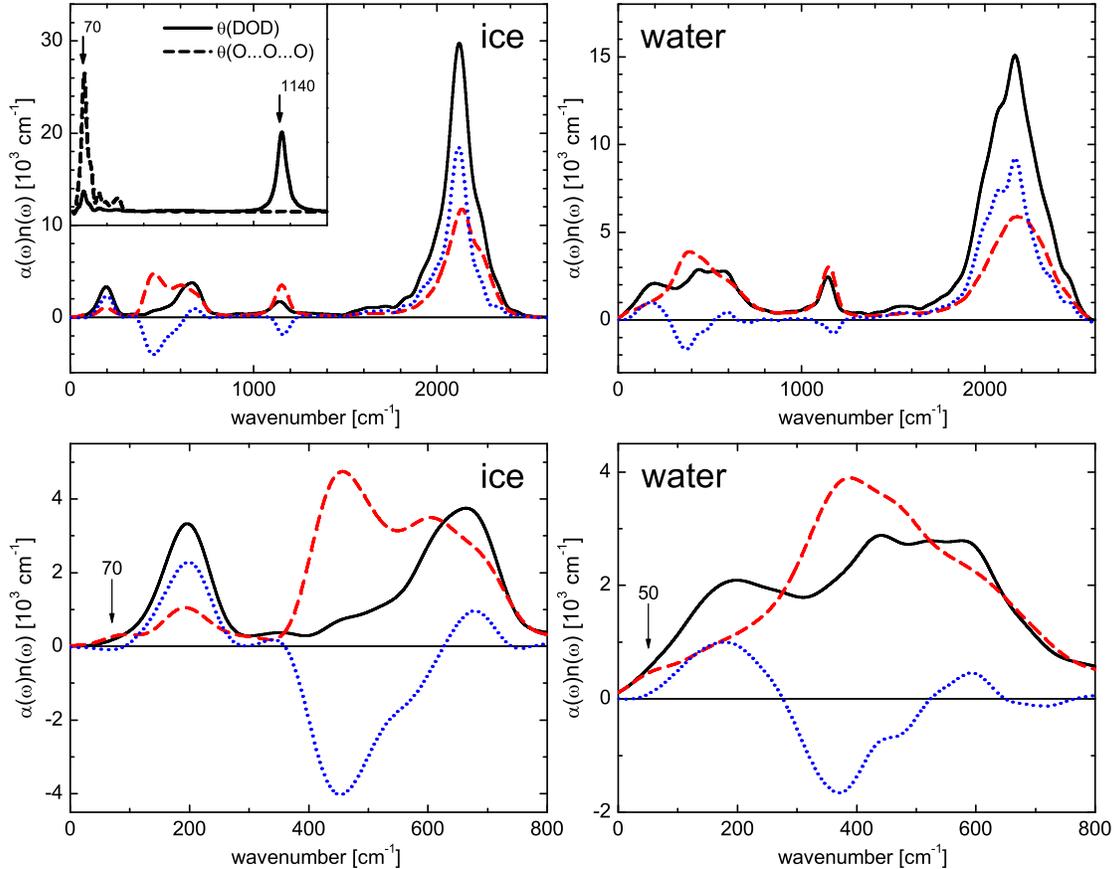}
\end{center}
	\caption{ 
		(color online). Intra (red dash) and inter (blue dot) contributions to the calculated IR spectra (black solid). Lower panels magnify the (0, 800) cm$^{-1}$ region. The inset shows the power spectrum of the autocorrelation function of the molecular angle $\theta$(DOD), and of the H-bond angle $\theta$(O...O...O), in arbitrary units. The arrows indicate positions in cm$^{-1}$.
			}
\label{intra_inter}
\end{figure*}

While separating intra and intermolecular contributions would be difficult or even impossible in experiment, in theory it can be easily achieved due to the additive nature of two-body correlation functions. In Fig.~\ref{intra_inter} we report the intra and inter contributions to the IR spectra of ice and water. 
The total spectrum is computed from Eq.~(\ref{lineshapeHA}), in which $\langle\bm{M}(0)\bm{M}(t)\rangle = \langle \sum_{ij} \bm{\mu}_i(0) \bm{\mu}_j(t) \rangle$. The intra contribution corresponds to retaining only the terms with $i=j$, the remaining terms add up to the inter contribution.

In water, the most prominent band around 2100 cm$^{-1}$ is due to the overlap of the symmetric and asymmetric OD stretching modes together with the overtone of the bending mode~\cite{Eisenberg69,Whalley77}. With two maxima, the inter contribution has larger intensity and lower frequency compared to its intra counterpart. This gives rise to the three broad features and the asymmetric shape of the band observed in experiments. Similar effects occur in ice, where three main features have been experimentally identified in the stretching band~\cite{Bergren78,Buch99}. In this case, however, we can not identify distinct intermolecular features within our chosen Fourier filter~\cite{Details_smooth}. Two peaks resembling the experimental features appear in the inter contribution when we reduce the filtering but they are too closely separated compared to experiment. In contrast to the stretching modes, the inter contributions to the bending modes around 1140 cm$^{-1}$ are negative in both ice and water. This effect has a simple intuitive explanation: due to the electrostatic attraction between opposite charges, OD stretching modes are correlated between neighboring molecules whereas bending modes are anti-correlated. In addition, we find that the broad shape of the small combination band at $\sim$1550 cm$^{-1}$ in water ($\sim$1660 cm$^{-1}$ in ice) is mostly due to the inter contribution.

The far infrared region is more complex. Two features have been identified in the libration band of the Raman spectrum of the liquid, but only one mode was believed to be IR active on the basis of a simple tetrahedral model with C$_{2v}$ symmetry~\cite{Eisenberg69}. A more recent IR experiment, however, suggested that both modes are visible in the liquid~\cite{Zelsmann95}. These modes have their counterpart in ice and, indeed, we find two IR features at $\sim$460 cm$^{-1}$ and at $\sim$660 cm$^{-1}$ in our calculated spectrum. Both features have intra and inter contributions but, interestingly, the inter contribution at $\sim$460 cm$^{-1}$ is negative, largely cancelling its intra counterpart. Instead the inter contribution at 660 cm$^{-1}$ is small and positive, slightly enhancing the corresponding spectral feature. As a result, the ice libration band has the skewed shape observed in experiments (Fig.~\ref{ir_expt}). In the liquid, the situation is similar with the features shifted to $\sim$370 cm$^{-1}$ and $\sim$590 cm$^{-1}$. Broadening due to disorder makes the lineshape look more symmetric.

The features at $\sim$50 cm$^{-1}$ in water and at $\sim$200 cm$^{-1}$ in water and ice are H-bond network modes with bending and stretching character, respectively~\cite{Zelsmann95, Sharma05}. In Ref.~\cite{Sharma05}, the inter character of the 200 cm$^{-1}$ mode was emphasized. A more accurate analysis reveals, however, that both intra and inter contributions are present, with an inter/intra ratio of 2:1 in ice and of 1:1 in water. This analysis is facilitated in the case of ice, where the network modes are well separated from the libration band.

\subsection{\label{sec33}Identification of the network bending mode}
Interestingly, we also understand the origin of the feature at $\sim$50 cm$^{-1}$. Albeit weak, this feature has been identified in experiments particularly below room temperature~\cite{Zelsmann95}. In our simulation it appears as a weak intramolecular feature in both water and ice, where the peak is at $\sim$70 cm$^{-1}$. Its intra character is surprising because the feature originates from network bending modes, as shown by the power spectrum of the autocorrelation function of the angle between two adjacent H-bonds, i.e. the angle defined by the oxygen atoms of three molecules linked by H-bonds. The corresponding spectrum for ice is reported in the inset of Fig.~\ref{intra_inter} and is sharply peaked at $\sim$70 cm$^{-1}$. We also report in the same inset the power spectrum of the autocorrelation function of the angle between the two covalent OD bonds in a molecule. As expected this spectrum is sharply peaked at $\sim$1140 cm$^{-1}$ but, interestingly, it also exhibits a weaker feature at $\sim$70 cm$^{-1}$ due to the modulation of the molecular angle induced by the network bending modes. Thus, the absence of inter character in the IR spectrum derives from the negligible intermolecular charge fluctuation associated to network bending modes, while the intra contribution reflects the modulation of the molecular dipole moment induced by these modes.

\subsection{\label{sec34}Spatial extent of dynamic dipolar correlations}
To gain insight on the spatial extent of the intermolecular correlations, we extend the approach of Ref.~\cite{Sharma07} to the dynamic domain. We define the density of the intermolecular dipole correlation function as:
\begin{equation}
D_{inter}(r,t) = \frac{1}{4\pi r^2 \Delta  r} 
\langle \sum_{i=1,N} \sum_{j\neq i} \bm{\mu}_i(0) \bm{\mu}_j(t) \delta _{ij}(t) \rangle \label{D_inter}
\end{equation}
where $r$ is the distance between the center of mass of molecule $j$ at time $t$ and molecule $i$ at time $0$. $\delta _{ij}(t)=1$ if at time $t$ molecule $j$ is in spherical shell $(r,r+\Delta r)$ centered on molecule $i$ at time $0$, whereas $\delta _{ij}(t)=0$ otherwise. The corresponding integrated contribution to the IR spectrum up to distance $R$ is given by:
\begin{equation}
\alpha(\omega)n(\omega)|_{0\rightarrow {R}} ^{inter} = \frac{2\pi\omega^2\beta}{3cV} \int_{0}^R d\bm{r} \int_{-\infty}^\infty 
dt e^{-i\omega t} D_{inter}(r,t)  \label{lineshape_inter}
\end{equation}

\begin{figure}[htp]
   \resizebox{\columnwidth}{!}{\includegraphics{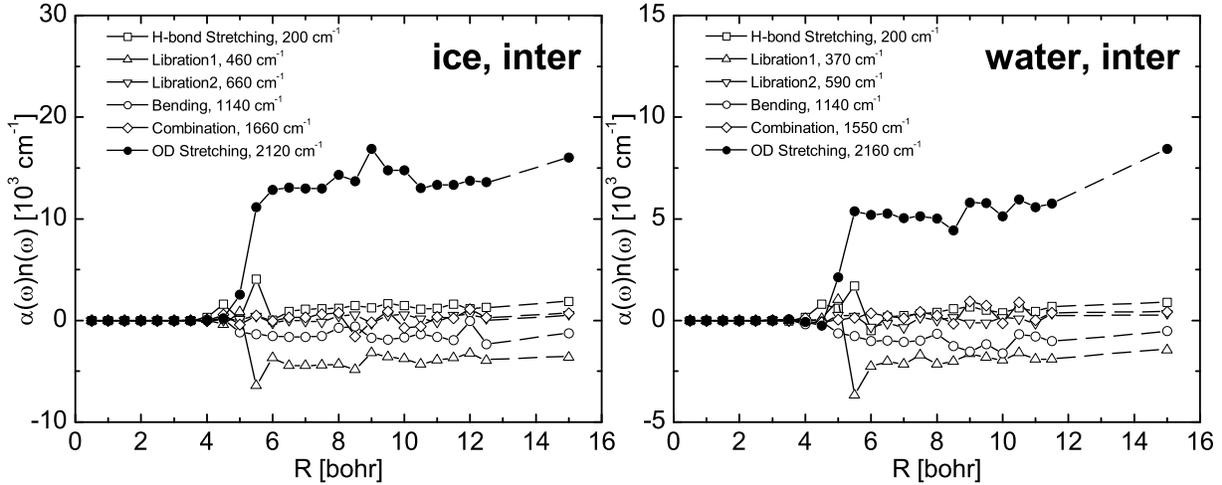}}    
   \caption{ 
            Inter contributions of the peak intensities vs. $R$ (Eq.~(\ref{lineshape_inter})). The maximum value of $R$ corresponds to half the size of the simulation cell. The points at 15 bohr report the peak values from Fig.~\ref{intra_inter}, which have been calculated without spherical cutoff.
           }
   \label{ir_w_r}
\end{figure}

The $R$ dependence of several peak intensities is reported in Fig.~\ref{ir_w_r}.
The curves indicate that the most important intermolecular correlations for all the modes occur when $R$ is comprised between 4 and 6 bohr, i.e. in correspondence with the first coordination shell, in both water and ice. Beyond 6 bohr only minor additional contributions are present, similar to the findings for static correlations in Ref.~\cite{Sharma07}. Short-range dynamic correlations are a direct manifestation of the presence of H-bonds between adjacent molecules. While static dipolar correlations due to H-bonds always enhance the dielectric response, dynamic correlations do enhance the system response at some frequencies but they also suppress it at other frequencies.

\section{\label{sec4}Conclusion}
Given the importance of intermolecular correlations, only models that include their effect at a sufficient level of detail can realistically describe the observed experimental features. An important result of our study is that these correlations play an equally important role over the entire spectral range and are not limited to the network modes that characterize the far IR range. We expect that similar dynamic correlations should also be crucial to interpret other spectroscopic data, such as Raman, which probe the adiabatic dynamics of the electrons. It would be difficult to include properly these effects in simulations based on classical empirical potentials. Our analysis shows that liquid water and ice are systems in which the molecules are strongly correlated, suggesting that their IR spectra cannot be simply explained in terms of a weakly perturbed collection of individual molecules. Finally, the decomposition technique introduced in this paper is not restricted to bulk water and ice, but can also be applied to other systems and to water at interfaces. In the latter case one may expect that characteristic signatures of hydrophilic and hydrophobic interfaces should emerge from this analysis.

\begin{acknowledgments}
We would like to thank Davide Donadio for useful discussions.
W.C. and R.C. gratefully acknowledge support from the NSF-MRSEC grant DMR-0213706.
M.S. and G.G. gratefully acknowledge support from Scidac grant No. DE-FG02-06ER46262.
\end{acknowledgments}


\begin{thebibliography}{31}
\bibitem{Details_Cutoff}
Increasing the plane wave cutoff in the water simulation to 85 Ry would not change our qualitative results. The main effect would be variations of a few percent in the peak frequencies.
\bibitem{Details_smooth}
In the calculated IR spectra in Fig.~\ref{ir_expt} and \ref{intra_inter} a Fourier filter is used to eliminate fluctuations with a width smaller than 60 cm$^{-1}$.
\end{thebibliography}
\end{document}